\documentclass[3p,times]{elsarticle}

\usepackage{amsmath}
\usepackage{amssymb}
\usepackage{graphicx}
\usepackage{enumitem} 
\newtheorem{Theorem}{Theorem}[section]
\newtheorem{Proposition}{Proposition}[section]
\newtheorem{Lemma}{Lemma}[section]

\newtheorem{Example}{Example}
\newtheorem{Definition}{Definition}[section]
\newenvironment{Proof}{\paragraph{Proof:}}{\hfill$\blacksquare$}
\newcommand{\norm}[1]{\left\lVert#1\right\rVert}

\usepackage{amssymb}

\usepackage[figuresright]{rotating}

\begin{document}

\begin{frontmatter}





\title{A weighted finite difference method for subdiffusive Black Scholes Model}
\author[label1]{Grzegorz Krzy\.zanowski}
\ead{grzegorz.krzyzanowski@pwr.edu.pl}

\author[label1]{Marcin Magdziarz}
\ead{marcin.magdziarz@pwr.wroc.pl}

\author[label1]{\L{}ukasz P\l{}ociniczak}
\ead{lukasz.plociniczak@pwr.edu.pl}

\address[label1]{Hugo Steinhaus Center,
Faculty of Pure and Applied Mathematics, Wroclaw University of Science and Technology
50-370 Wroclaw, Poland}

\begin{abstract}
In this paper we focus on the subdiffusive Black-Scholes (B-S) model.
The main part of our work consists of the finite difference method as a numerical approach to the option pricing in the considered model. We find the governing fractional differential
equation and the related weighted numerical scheme being a generalization of the classical Crank-Nicolson (C-N) scheme.
The proposed method has $2-\alpha$ order of accuracy with respect to time where $\alpha\in(0,1)$ is the subdiffusion parameter, and $2$ with respect to space. Further, we provide the stability and convergence analysis.
Finally, we present some numerical results. 
\end{abstract}


\end{frontmatter}

\section{Introduction}
\label{intro}
Options are one of the most popular and important financial derivatives, therefore the question about their valuation
has an essential meaning for financial institutions and global economy.
The celebrated B-S formula for European options \cite{Black,Meerton} was of such great importance that
the authors were awarded the Nobel Prize in Economics in 1997. After recent investigations \cite{Sch} it seems that the B-S-Merton
formula, despite its simplicity and clarity, cannot be used in many cases. 
 One of the most vivid examples is the case when the dynamics of the underlying instrument has a tendency 
 to have constant periods or sudden jumps \cite{jumps}. The feature of the market stagnation can be observed e.g. in emerging markets \cite{bor} and in interest rate behavior \cite{bdt}. The classical B-S
model was proposed under some strict assumptions, however some improved models have been established to
weaken these assumptions, such as stochastic volatility model \cite{Hull}, stochastic interest model
\cite{Merton},
models with transactions costs \cite{Barles,Davis}, jump - diffusion model \cite{Merton2}. 

In recent years the theory of fractional differential equations found important applications in econometrics and finance \cite{sikorski}.
Also many researchers have investigated the generalization of the B-S equation into fractional case. 
The reason of this generalization is the fractal structure for financial market and increasing interest of fractional calculus.
Usually the procedure is to replace the
standard Brownian motion used in the classical model
by the fractional Brownian motion \cite{H6,H8,H7}. More precisely, the order of time
variation $dt$ is replaced by the Hurst exponent $H$ $(0<H<1)$. 
Changing the parameter of self-similarity out of the case $H=1/2$ leads to the lack of martingale property of the process describing dynamics 
of financial asset. It is equivalent that such generalized model allows arbitrage.
We proceed with a completely different approach. We 
replace the Geometric Brownian Motion $Z(t)$ by the subdiffusive Geometric Brownian Motion $Z(S_{\alpha}(t))$
with the inverse subordinator $S_{\alpha}(t)$ \cite{sato} in the definition of the process describing the underlying asset. In this way we find the corresponding
fractional differential equation, which is different than most of considered in the literature but the same as is given in \cite{udin} and \cite{zhang}.
For other results related to subdiffusive B-S model see \cite{gu2, chin11, janwyl, wyss}.

 Many efficient numerical methods have been proposed for solving fractional differential equations,
 which include finite difference methods, finite element methods,
 finite volume methods,
spectral methods and meshless methods (see \cite{zhang} and references therein).

Developing numerical discretization methods for fractional integrals
and derivatives is one of the important topics in fractional calculus due to its wide applications. 
In this work we propose the quadrature method for approximation the Caputo derivative 
which implies the order of accuracy equal to $2-\alpha$ . In \cite{Gao,Cap} authors studied approximations of order $3-\alpha$, while in 
\cite{Cao} of $4-\alpha$. In \cite{Cap} the Caputo derivative was approximated using the
$r+1$-th Lagrange interpolation, and obtained a series of high-order numerical
schemes with accuracy of $r+1-\alpha$ at $n$-th steps ($n\geq r$). With higher order of accuracy the level of complexity increases
and the stability of the numerical scheme can be lost. 
\section{Subdiffusive B-S Model} 
\subsection{Assumptions of the subdiffusive B-S Model}
Let us consider a market, whose evolution is taking place up to time horizon $T$ and is contained in the probability space 
$(\Omega, \mathcal{F} , \mathbb{P})$.
Here, $\Omega$ is the sample space,  $\mathcal{F}$ is filtration interpreted as the information about history of asset price which completely is available for the
investor and $\mathbb{P}$ is the “objective” probability measure. The assumptions are the same as in  the classical case \cite{ja} with the exception that we do not have to assume the market liquidity and that 
the underlying instrument instead of Geometric Brownian Motion (GBM) has to follow subdiffusive GBM \cite{MM}:
$$
\left\{ \begin{array}{ll}
Z_{\alpha}\left(t\right)=Z\left(S_{\alpha}(t)\right),\\
Z\left(0\right)=Z_{0},                                                         
\end{array}\right.$$
where $Z_{\alpha}\left(t\right)$ - the price of the underlying asset, $Z(t)=Z\left(0\right)\exp \left(\mu t+\sigma B_{t}\right)$, $\sigma$ - volatility (constant), $\mu$ - drift (constant), $B_{t}$ - Brownian motion, $S_{\alpha}(t)$ - the inverse $\alpha$-stable subordinator defined as $S_{\alpha}(t)=\inf(\tau>0: U_{\alpha}(\tau)>t)$ \cite{MM}, where $U_{\alpha}(t)$ is the $\alpha$-stable subordinator \cite{sato}, $0<\alpha<1$. We assume $S_{\alpha}(t)$ is independent  of $B_{t}$ for each $t\in[0,T]$.

\begin{figure}[h]
\raggedleft
\includegraphics[scale=0.35]{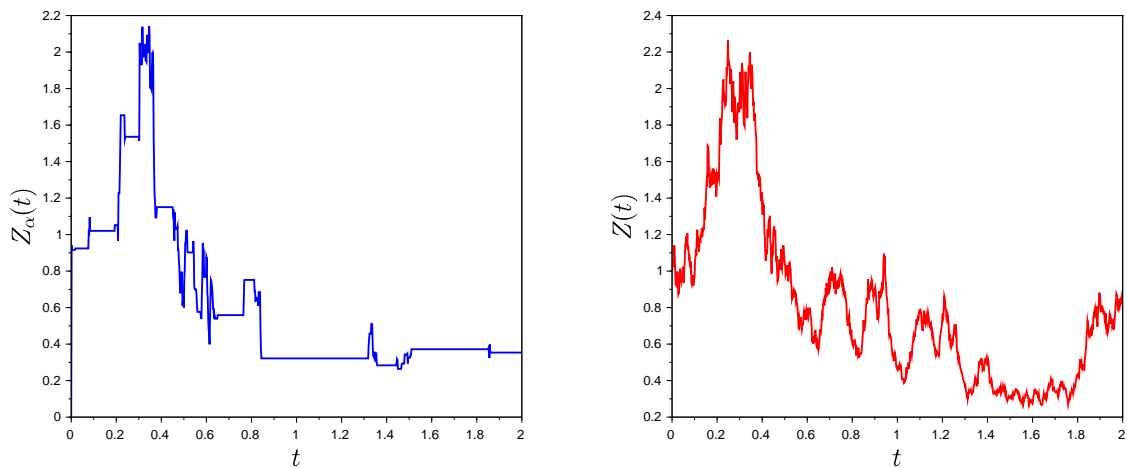}
\caption{The sample trajectory of subdiffusive GBM (left) with its classical analogue (right). In the subdiffusive case the constant periods characteristic for
emerging markets can be observed. The parameters are $Z_{0}=\sigma=\mu=1$, $\alpha=0.7$.}   
\end{figure}

The subdiffusive B-S is the generalization of the standard model to the cases, where the markets can be non-liquid. Then the subdiffusion parameter $\alpha\in(0,1)$ can be considered as the measure of non-liquidness. The constant periods appear more frequently with decreasing $\alpha$. With $\alpha\rightarrow 1$ the subdiffusive B-S model reduces to the classical (liquid) case. Due to its simplicity and practicality, the classical B-S Model
is one of the most widely used in option pricing. Although in contrast to the subdiffusive case it does not take into account the empirical property of constant price periods.
In Figure $1$ we compare sample simulation of underlying asset in classical and subdiffusive market model. Even short stagnation of a market can not be simulated by classical B-S model.
As a generalization of the B-S model, its subdiffusive equivalent can be used in wide range cases - including all markets where B-S can be applied.
The method of subdiffusive B-S model calibration from empirical data is described in \cite{orzel}.

By the classical put-call parity \cite{musiela} we have following fact:
\begin{Proposition}\cite{MM}
  For the fair price of European call and put options in subdiffusive B-S model we have following relationship:
  \[C_{BS}^{sub}\left(Z_{0},K,T,\sigma,r, \alpha\right)-P_{BS}^{sub}\left(Z_{0},K,T,\sigma, r, \alpha\right)=Z_{0}-Ke^{-rT}.\]
 \end{Proposition}
Here and in whole paper: $C_{BS}^{sub}$ - fair price of European call, $P_{BS}^{sub}$ - fair price of European put, $K$ - strike,
$T$ - maturity,
$\sigma$ - volatility,
$r$ - interest rate.\\
 \\ One of the most expected property of the market is that there is no possibility to gain money without taking the risk. This property is called 
 the lack of arbitrage and formally means that the self-financing strategy $\phi$ which follows to a positive profit without any probability of intermediate loss can not be constructed \cite{4}.
 By the Fundamental theorem of asset pricing \cite{4}, the market model described by $(\Omega, \mathcal{F}, \mathbb{P})$ and
underlying instrument $Z_{\alpha}(t)$ with filtration $\mathcal{F}_{t\in[0,T]}$ is arbitrage-free if and only if there exists
a probability measure $\mathbb{Q}$, (called the risk-neutral measure) equivalent to $\mathbb{P}$ such that the asset $Z_{\alpha}(t)$ is a martingale
with respect to $\mathbb{Q}$. Under this measure,
financial instruments have the same expected rate of return, regardless the variability of the prices.
This is in contrast to the physical probability measure (the actual probability
distribution of prices), under which more risky instruments have a higher expected rate of return
than less risky instruments.

  Let us introduce the probability measure \begin{equation}
                                            \mathbb{Q}(A)=\int_{A}\exp\left(-\gamma B(S_{\alpha}(T))-\frac{\gamma^{2}}{2}S_{\alpha}(T)\right)dP,
                                           \end{equation}
  where $\gamma=\frac{\mu+\frac{\sigma^2}{2}}{\sigma}$, $A\in\mathcal{F}$.
  As it is shown in \cite{MM} the process $Z_{\alpha}(t)$ is martingale with respect to $\mathbb{Q}$, so we have the following
 
 \begin{Theorem}\cite{MM}
  The subdiffusive B-S Model is arbitrage-free.
 \end{Theorem}

Another property of market model is the so-called completeness. Intuitively the market model is complete
if the set of possible gambles on future states-of-the-world can
be constructed with existing assets. More formally, the market model is complete if every
$\mathcal{F}_{t\in[0,T]}$ -measurable random variable $X$ admits a replicating self-
financing strategy $\phi$ \cite{4}.
The Second Fundamental theorem of asset pricing \cite{4} states that a
market model described by $(\Omega, \mathcal{F}, \mathbb{P})$ and
underlying instrument $Z_{\alpha}(t)$ with filtration $\mathcal{F}_{t\in[0,T]}$ is complete if and only if there is a unique
martingale measure equivalent to $\mathbb{P}$. 
 \begin{Theorem}\cite{MM}
  The market model in which the price of underlying instrument follows the subdiffusive GBM $Z_{\alpha}(t)$
is incomplete.
 \end{Theorem}

Market incompleteness means that there is no unique fair
price of financial derivatives, because for different martingale measures different prices could be obtained. 
Despite $\mathbb{Q}$ defined in $(1)$ is not unique, in the sense of criterion of minimal relative entropy it is the ``best'' martingale
measure. 
It means that the measure $\mathbb{Q}$ minimizes the distance to the
measure $\mathbb{P}$ \cite{Gajda} . Other essential fact is that for $\alpha\to1$, $\mathbb{Q}$ reduces to the measure of the classical
B-S model which is arbitrage-free and complete. It is consistent with our intuition if we consider subdiffusive B-S model as the generalization of the standard B-S model.
Thus, in whole paper we will use the martingale measure $\mathbb{Q}$ defined in $(1)$ as a reference measure.

\subsection{The fair price of a call option in the subdiffusive B-S model}
Let us define the fair price of a call option for subdiffusive and classical B-S model:
 \begin{equation*}
 v\left(t\right)=C_{BS}^{sub}\left(Z_{0},K,t,r,\sigma,\alpha \right),
\end{equation*}
 \begin{equation*}h\left(t\right)=C_{BS}\left(Z_{0},K,t,r,\sigma \right).\end{equation*}
By the B-S formula \cite{Black,Meerton}, we have  \begin{equation*}h\left(t\right)=Z_{0}\Phi(d_{+})-Ke^{-rt}\Phi(d_{-}),\end{equation*}
with $\displaystyle d_{\pm}=\frac{\log \frac{Z_{0}}{K}+\left(r\pm\frac{1}{2}\sigma^{2}\right)t}{\sigma\sqrt{t}}.$ Here $\Phi$ denotes the cumulative distribution function of standard normal distribution.\\
By \cite{MM}, the relation between functions $v$ and $h$ is as follow 
\begin{equation*}
 v\left(t\right)=\mathbf{E} h(S_{\alpha}(t))=\int_{0}^{\infty}h\left(z\right)\varrho_{\alpha}\left(z,t\right)dz,
\end{equation*}
where, $\varrho_{\alpha}(x,t)$ is the PDF of $S_{\alpha}(t)$.
Moreover, $v(t)$ follows
\begin{equation*}
 v\left(t\right)=\int_{0}^{\infty}h(x)t^{-\alpha}g_{\alpha}(x/t^{\alpha})dx,
\end{equation*}
where $g_{\alpha}(z)$ is given in terms of Fox function $H^{1,0}_{1,1}(z^{(1-\alpha,\alpha)}_{(0,1)})$ (see \cite{MM} and references therein).

Note that  \begin{equation*}v\left(0\right)=h\left(0\right).\end{equation*}
Let us consider the Laplace transform of the function $v$:\begin{equation*}
\mathcal{L}\{v\}(k)=\displaystyle\int_{0}^{\infty}e^{-kt}v\left(t\right)dt=\displaystyle\int_{0}^{\infty}e^{-kt}\displaystyle\int_{0}^{\infty}h\left(z\right)\varrho_{\alpha}\left(z,t\right)dzdt=
 \displaystyle\int_{0}^{\infty}\int_{0}^{\infty}h\left(z\right)\varrho_{\alpha}\left(z,t\right)e^{-kt}dtdz=\displaystyle\int_{0}^{\infty}h\left(z\right)\mathcal{L}\{\varrho_{\alpha}\}\left(z,k\right)dz.
 \end{equation*}
 In the same way as in \cite{MM} we find that $\mathcal{L}\{\varrho_{\alpha}\}=k^{\alpha-1}e^{-zk^{\alpha}}$. Hence, we have:
 \begin{equation*}
 \displaystyle\int_{0}^{\infty}h\left(z\right)\mathcal{L}\{\varrho_{\alpha}\}\left(z,k\right)dz=\int_{0}^{\infty}h\left(z\right)k^{\alpha-1}e^{-zk^{\alpha}}dz
=k^{\alpha-1}\int_{0}^{\infty}h\left(z\right)e^{-zk^\alpha}dz=k^{\alpha-1}\mathcal{L}\{h\}\left(k^{\alpha}\right).\end{equation*}
So as a conclusion we have the following result:
\begin{equation}
\mathcal{L}\{h\}\left(k\right)=k^{(1-\alpha)/\alpha}\mathcal{L}\{v\}\left(k^{1/\alpha}\right).
\end{equation}
Let us write the B-S equation describing $h\left(z,t\right)$ \cite{ja}:
$$
\left\{ \begin{array}{ll}
  \displaystyle\frac{\partial h\left(z,t\right)}{\partial t}+\frac{1}{2}\sigma^{2}z^{2}\frac{\partial^2 h\left(z,t\right)}{\partial z^2}+rz\frac{\partial h\left(z,t\right)}{\partial z}-rh\left(z,t\right)=0,\\       
       h\left(z,T\right)=\max(z-K,0).
       \end{array}\right.
       $$
       Now let us take the Laplace transform with respect to $t$:
       \[k\mathcal{L}\{h\}\left(k\right)-h\left(0\right)+\frac{1}{2}\sigma^2z^2\frac{\partial^2 \mathcal{L}\{h\}\left(k\right)}{\partial z^2}+rz\frac{\partial \mathcal{L}\{h\}\left(k\right)}{\partial z}-r\mathcal{L}\{h\}\left(k\right)=0,\]
       Then we use formula $(2)$ and the fact that $v(0)=h(0)$, obtaining:
            \[\mathcal{L}\{v\}\left(k^{\frac{1}{\alpha}}\right)k^{\frac{1}{\alpha}}-v\left(0\right)+\frac{1}{2}\sigma^2z^{2}k^{\frac{1-\alpha}{\alpha}}\frac{\partial^2 \mathcal{L}\{v\}\left(k^{\frac{1}{\alpha}}\right)}{\partial z^2}+rzk^{\frac{1-\alpha}{\alpha}}\frac{\partial \mathcal{L}\{v\}\left(k^{\frac{1}{\alpha}}\right)}{\partial z}-k^{\frac{1-\alpha}{\alpha}}r\mathcal{L}\{v\}\left(k^{\frac{1}{\alpha}}\right)=0.\]
Now let us change variable - we replace $k$ by $k^\alpha$:            
            \[k\mathcal{L}\{v\}\left(k\right)-v\left(0\right)=
            k^{1-\alpha}\left(-\frac{1}{2}\sigma^2z^{2}
            \frac{\partial^2 \mathcal{L}\{v\}\left(k\right)}{\partial z^2}-rz\frac{\partial \mathcal{L}\{v\}\left(k\right)}{\partial z}
            +r\mathcal{L}\{v\}\left(k\right)\right).\]
Inverting the Laplace transform, we get:
\begin{equation}\frac{\partial v\left(z,t\right)}{\partial t}= {}_{0}^{RL}D_{t}^{1-\alpha}\left(-\frac{1}{2}\sigma^2z^{2}\frac{\partial^2 v\left(z,t\right)}{\partial z^2}-rz\frac{\partial v\left(z,t\right)}{\partial z}+rv\left(z,t\right)\right),
 \end{equation}

where $\alpha\in\left(0,1\right)$, ${}_{0}^{RL}D_{t}^{\alpha}$ is Riemann-Louville fractional derivative defined as
\[{}_{0}^{RL}D_{t}^{\alpha}g\left(t\right)=\frac{1}{\Gamma\left(1-\alpha\right)}\frac{d}{dt}\int_{0}^{t}\left(t-s\right)^{-\alpha}g\left(s\right)ds.\]
Above we used the fact that for $\alpha\in\left(0,1\right)$, the Laplace transform for the Riemann-Louville fractional derivative follows \cite{podlubny}: 
   \begin{equation*}\displaystyle \mathcal{L}\{{}_{0}^{RL}D_{t}^{\alpha}f\left(t\right)\}=k^{\alpha}\mathcal{L}\{f\}(k)- {}_{0}^{RL}D_{t}^{\alpha-1}f\left(t\right)_{t=0},
   \end{equation*}
where $f\in C^{1}$. Note that the Laplace transform for the Caputo fractional derivative follows \cite{podlubny}:
 \begin{equation*}\mathcal{L}\{{}_{0}^{c}D_{t}^{\alpha}f\left(t\right)\}=k^{\alpha}\mathcal{L}\{f\}(k)-\frac{f(0)}{k^{1-\alpha}},
   \end{equation*}
where $f\in C^{1}$, $\alpha\in\left(0,1\right)$ and ${}_{0}^{c}D_{t}^{\alpha}$ is defined as \cite{podlubny}
\[{}_{0}^{c}D_{t}^{\alpha}g\left(t\right)=\frac{1}{\Gamma\left(1-\alpha\right)}\int_{0}^{t}\frac{d g\left(s\right)}{d s}\left(t-s\right)^{-\alpha}ds.\]
Using basic properties of fractional derivatives, we transform $(3)$  into
\[{}_{0}^{c}D_{t}^{\alpha}v\left(z,t\right)= -\frac{1}{2}\sigma^2z^{2}\frac{\partial^2 v\left(z,t\right)}{\partial z^2}-rz\frac{\partial v\left(z,t\right)}{\partial z}+rv\left(z,t\right).\]
By this way we have found the following system:
\begin{equation*}
 \begin{cases}
 \displaystyle {}_{0}^{c}D_{t}^{\alpha}v\left(z,t\right)=
 -\frac{1}{2}\sigma^2z^{2}\frac{\partial^2 v\left(z,t\right)}{\partial z^2}-rz\frac{\partial v\left(z,t\right)}{\partial z}+rv\left(z,t\right),\\
v\left(z,T\right)=\max\left(z-K,0\right),\\
 v\left(0,t\right)=0,\\
\lim\limits_{z\rightarrow \infty}v\left(z,t\right)\sim z, 
\end{cases}
\end{equation*}
  for $\left(z,t\right)\in\left(0,\infty\right)\times\left(0,T\right)$.

Let us introduce the following variable:
\begin{equation}
 x=\ln z
\end{equation}

and the function:
\begin{equation}
u\left(x,t \right)=v\left(e^{x},T-t \right).
\end{equation}
Hence, we have:
\begin{Theorem}
The fair price of a call option in the subdiffusive B-S model is equal to $v(z,t)$, where $v(z,t)$ satisfies $(4)$ and $(5)$, and $u(x,t)$ is 
the solution of the system:
\begin{equation}
 \begin{cases}
{}_{0}^{c}D_{t}^{\alpha}u\left(x,t\right)= \displaystyle\frac{1}{2}\sigma^2\displaystyle\frac{\partial^2 u\left(x,t\right)}{\partial x^2}+\left(r-\displaystyle\frac{1}{2}\sigma^2\right)\displaystyle\frac{\partial u\left(x,t\right)}{\partial x}-ru\left(x,t\right),\\
u\left(x,0\right)=\max\left(e^x-K,0\right),\\
\lim\limits_{x\rightarrow -\infty} u\left(x,t\right)=0,\\
\lim\limits_{x\rightarrow \infty}u\left(x,t\right)\sim e^{x},\\
\end{cases}
\end{equation}
for $\left(x,t\right)\in\left(-\infty,\infty\right)\times\left(0,T\right)$.
\end{Theorem}
The solution of $(6)$ exists and it is unique (see \cite{exist} and references therein).
\section{Finite difference method}
To solve the above model numerically we will approximate limits by finite numbers and derivatives by finite differences. After obtaining the discrete analogue of $(6)$ we will solve the problem recursively using initial-boundary conditions. We will proceed similarly, as it was done for implicit method in \cite{zhang}.
\subsection{Weighted scheme for the subdiffusive B-S model}
The system $(6)$ has the following form:
\begin{equation*}
 \begin{cases}
{}_{0}^{c}D_{t}^{\alpha}u\left(x,t\right)= a\displaystyle\frac{\partial^2 u\left(x,t\right)}{\partial x^2}+b\displaystyle\frac{\partial u\left(x,t\right)}{\partial x}-cu\left(x,t\right),\\
u\left(x,0\right)=f\left(x\right),\\
u\left(x_{min},t\right)=p\left(t\right),\\
u\left(x_{max},t\right)=q\left(t\right),\\
\end{cases}
\end{equation*}
where $a=\displaystyle\frac{1}{2}\sigma^2$, $b=\left(r-\displaystyle\frac{1}{2}\sigma^2\right)$, $c=r$, $\displaystyle f\left(x\right)=\max\left(\exp(x)-K,0\right)$, $\displaystyle p\left(t\right)\rightarrow 0$ if $\displaystyle x_{min}\rightarrow -\infty$,
$\displaystyle q\left(t\right)\rightarrow \exp(x_{max})$ if $\displaystyle x_{max}\rightarrow \infty$. The put-call parity implies that $\displaystyle p\left(t\right)=0$, $\displaystyle q\left(t\right)=\exp(x_{max})-K\exp(-r\left(T-t\right))$.  

Let us denote \[b_{j}=\left(j+1\right)^{1-\alpha}-j^{1-\alpha},\]
\[d=\Gamma\left(2-\alpha\right)\Delta t^{\alpha},\]
\[u^{k}=\left(u^{k}_{1},u^{k}_{2},\dots, u^{k}_{n-1}\right)^{T},\]
\[G^{k}=\left(\left(\displaystyle\frac{ad}{\Delta x^2}-\displaystyle\frac{bd}{2\Delta x}\right)u^{k}_{0},0,\dots ,0,\left(\displaystyle\frac{ad}{\Delta x^2}+\displaystyle\frac{bd}{2\Delta x}\right)u^{k}_{n}\right)^{T},\]
\[f=\left(0,f_{2},f_{3},\dots f_{n-3},f_{n-2},q^{0}\right)^{T},\]
where $u^{k}_{i}=u\left(x_{i},t_{k}\right)$, $f_{i}=f\left(x_{i}\right)$, $i=2,\dots, n-2$, $q^{k}=q\left(t_{k}\right)$, $k=0,1,\dots,N$, moreover $\displaystyle  \Delta t=T/N$, $\displaystyle  \Delta x=(x_{max}-x_{min})/n$ are time and space steps respectively. 
We will use the following approximations for space derivatives:
\begin{equation*}
\begin{cases}
 \displaystyle\frac{\partial u\left(x_{i},t_{k+1}\right)}{\partial x}=\displaystyle\frac{u\left(x_{i+1},t_{k+1}\right)-u\left(x_{i-1},t_{k+1}\right)}{2\Delta x}+O\left(\Delta x^{2}\right),\\
\displaystyle\frac{\partial^{2} u\left(x_{i},t_{k+1}\right)}{\partial x^{2}}=\displaystyle\frac{u\left(x_{i+1},t_{k+1}\right)-2u\left(x_{i},t_{k+1}\right)+u\left(x_{i-1},t_{k+1}\right)}{\Delta x^{2}}+O\left(\Delta x^{2}\right).
\end{cases}
\end{equation*}
We approximate the fractional-time derivative by:
\[{}_{0}^{c}D_{t}^{\alpha}u\left(x_{i},t_{k}\right)=
\displaystyle\frac{1}{\Gamma \left(2-\alpha \right)}\displaystyle\sum_{j=0}^{k}\displaystyle\frac{u\left(x_{i},t_{k+1-j}\right)-u\left(x_{i},t_{k-j}\right)}{\Delta t^{\alpha}}\left(\left(j+1\right)^{1-\alpha}-j^{1-\alpha}\right)+O\left(\Delta t^{2-\alpha}\right).\]
After omitting the truncation errors, the implicit discrete scheme 
can be expressed in following form: 

\begin{equation}
\begin{cases}
A\hat{u}^{1}=\hat{u}^{0}+G^{1},\\
A\hat{u}^{k+1}=\displaystyle\sum_{j=0}^{k-1}\left(b_{j}-b_{j+1}\right)\hat{u}^{k-j}+b_{k}\hat{u}^{0}+G^{k+1},
\end{cases}
\end{equation}
where $k\geq1$, $A=\left(a_{ij}\right)_{\left(n-1\right)\times\left(n-1\right)}$, such  that:\newline
$a_{ij}=$
 $
\begin{cases}
1+2\displaystyle\frac{ad}{\Delta x^2}+cd,& \text{for } j=i,\text{ }i=1,2\dots ,n-1,\\
-\left(\displaystyle\frac{ad}{\Delta x^2}-\displaystyle\frac{bd}{2\Delta x}\right),& \text{for } j=i-1,\text{ }i=2\dots ,n-1,\\
-\left(\displaystyle\frac{ad}{\Delta x^2}+\displaystyle\frac{bd}{2\Delta x}\right),& \text{for } j=i+1,\text{ }i=1\dots ,n-2,\\
0,& \text{in other cases}
\end{cases}
$\\
\\The corresponding initial-boundary conditions are
\begin{equation}
\begin{cases}
\hat{u}^{0}=f,\\
\hat{u}^{k}_{0}=0,\\
\hat{u}^{k}_{n}=q^{k},
\end{cases}
\end{equation}
where $k\geq1$. In similar way let us write the explicit discrete scheme. We use approximations for space derivatives as follows:
\begin{equation}
\begin{cases}
\displaystyle\frac{\partial u\left(x_{i},t_{k}\right)}{\partial x}=\displaystyle\frac{u\left(x_{i+1},t_{k}\right)-u\left(x_{i-1},t_{k}\right)}{2\Delta x}+O\left(\Delta x^{2}\right),\\
\displaystyle\frac{\partial^{2} u\left(x_{i},t_{k}\right)}{\partial x^{2}}=\displaystyle\frac{u\left(x_{i+1},t_{k}\right)-2u\left(x_{i},t_{k}\right)+u\left(x_{i-1},t_{k}\right)}{\Delta x^{2}}+O\left(\Delta x^{2}\right).
\end{cases}
\end{equation}
In the matrix form the explicit discrete scheme
can be expressed in the following form:
\begin{equation}
\begin{cases}
\hat{u}^{1}=B\hat{u}^{0}+\hat{u}^{0}+G^{0},\\
\hat{u}^{k+1}=B\hat{u}^{k}+\displaystyle\sum_{j=0}^{k-1}\left(b_{j}-b_{j+1}\right)\hat{u}^{k-j}+ b_{k}\hat{u}^{0}+G^{k},
\end{cases}
\end{equation}
where $k\geq1$, $B=\left(b_{ij}\right)_{\left(n-1\right)\times\left(n-1\right)}$, such  that:\newline
$b_{ij}=$
 $
\begin{cases}
-\left(2\displaystyle\frac{ad}{\Delta x^2}+cd\right),& \text{for } j=i,\text{ }i=1\dots ,n-1,\\
\displaystyle\frac{ad}{\Delta x^2}+\displaystyle\frac{bd}{2\Delta x},& \text{for } j=i+1,\text{ }i=1\dots ,n-2,\\
\displaystyle\frac{ad}{\Delta x^2}-\displaystyle\frac{bd}{2\Delta x},& \text{for } j=i-1,\text{ }i=2\dots ,n-1,\\
0,& \text{in other cases}
\end{cases}
$\\
\\Taking the linear combination of $(7)$ and $(10)$ we obtain a weighted scheme:

\begin{equation}
\begin{cases}
C\hat{u}^{1}=\hat{u}^{0}+\left(1-\theta\right)G^{1} +\theta G^{0}+\theta B\hat{u}^{0},\\
C\hat{u}^{k+1}=\displaystyle\sum_{j=0}^{k-1}\left(b_{j}-b_{j+1}\right)\hat{u}^{k-j}+b_{k}\hat{u}^{0}+\left(1-\theta\right)G^{k+1} +\theta G^{k}+\theta B\hat{u}^{k},
\end{cases}
\end{equation}
where $k\geq1$, $C=\theta I+\left(1-\theta\right) A$, $\theta\in[0,1]$ 
and the corresponding initial-boundary conditions are defined in $(8)$.
Let us denote that in the case of the classical B-S model $\theta=1/2$ defines the C-N scheme \cite{ja}. Motivated by this fact in whole paper we assume the following:
\begin{Definition}
 Scheme $(11)$ with $\theta=1/2$ is called the C-N discrete scheme.
\end{Definition}

\subsection{Consistency of the weighted discrete scheme}
In this section we will show the following 
\begin{Theorem}
 For $\theta\in[0,1]$ and $1\leq i\leq n,1\leq j\leq N$, the truncation error $R_{i}^{j}$ follows
\[\left|R_{i}^{j}\right|\leq C_{max}\Delta t^{\alpha}\left(\Delta t^{2-\alpha}+\Delta x^{2}\right).\]
\end{Theorem}

\begin{Proof}
As it was shown in \cite{Lin}, the Caputo derivative can be expressed as follows
\[\displaystyle{}_{0}^{c}D_{t}^{\alpha}=\frac{1}{\Gamma(1-\alpha)}\sum_{j=0}^{k}\frac{u(x,t_{k+1-j})-u(x,t_{k-j})}{\Delta t^{\alpha}}\left((j+1)^{1-\alpha}-j^{1-\alpha}\right)+r^{k+1}_{\Delta t},\]
where 
\[r^{k+1}_{\Delta t}\leq C_{u}\Delta t^{2-\alpha}.\]
On the other hand, we can apply $(9)$.
The full formulation of the discrete weighted scheme with the truncation error has the form:
\begin{equation}
\begin{cases}
-\left(\displaystyle\frac{ad}{\Delta x^{2}}+\displaystyle\frac{bd}{2\Delta x}\right)(\theta u^{0}_{i+1}+(1-\theta) u^{1}_{i+1})+\left(\displaystyle\frac{2ad}{\Delta x^{2}}+cd\right)(\theta u^{0}_{i}+(1-\theta) u^{1}_{i})-\\\left(\displaystyle\frac{ad}{\Delta x^{2}}-\displaystyle\frac{bd}{2\Delta x}\right)(\theta u^{0}_{i-1}+(1-\theta) u^{1}_{i-1})=  u_{i}^{0}-u_{i}^{1}+(1-\theta)R_{i}^{1}+\theta R_{i}^{0},\\
-\left(\displaystyle\frac{ad}{\Delta x^{2}}+\displaystyle\frac{bd}{2\Delta x}\right)(\theta u^{k}_{i+1}+(1-\theta) u^{k+1}_{i+1})+\left(\displaystyle\frac{2ad}{\Delta x^{2}}+cd\right)(\theta u^{k}_{i}+(1-\theta) u^{k+1}_{i})- \\\left(\displaystyle\frac{ad}{\Delta x^{2}}-\displaystyle\frac{bd}{2\Delta x}\right)(\theta u^{k}_{i-1}+(1-\theta) u^{k+1}_{i-1})=b_{k}u_{i}^{0}-u_{i}^{k+1}+
\displaystyle\sum_{j=0}^{k-1}\left(b_{j}-b_{j+1}\right)u_{i}^{k-j}\\+(1-\theta)R_{i}^{k+1}+\theta R_{i}^{k} .
\end{cases}
\end{equation}
Here $k\geq1$, $i=1,2,\dots ,n-1$ and the corresponding initial-boundary conditions are defined in $(8)$. 
By the approximation of the Caputo derivative and $(9)$ we have
\[\left|R_{i}^{j}\right|\leq C_{i}^{j}\Delta t^{\alpha}(\Delta t^{2-\alpha}+\Delta x^{2}),\]
where $C_{i}^{j}$ are constants ($1\leq i\leq n,1\leq j\leq N$).\\
Let us denote $\displaystyle C_{max}=\max_{1\leq i\leq n,1\leq j\leq N}C_{i}^{j}$.
Then, for the truncation error it holds that
\[\left|R_{i}^{j}\right|\leq C_{max}\Delta t^{\alpha}\left(\Delta t^{2-\alpha}+\Delta x^{2}\right).\]
Note that the parameter $\theta$ has no influence in the above analysis.
\end{Proof}

\subsection{Stability of the weighted discrete scheme}
We will proceed using von Neumann method.
For $l=0,1,\dots n$, $k=0,1,..,N$ let us denote
$u_{l}^{k}=u\left(x_{l},t_{k}\right)$ - the exact solution of numerical scheme,
$\hat{u}_{l}^{k}$ - some approximation of $u_{l}^{k}$. After omitting truncation error and introducing $e_{l}^{k}=u_{l}^{k}-\hat{u}_{l}^{k}$, $(12)$ has the form: 
\begin{equation}
\begin{cases}
-\left(\displaystyle\frac{ad}{\Delta x^{2}}+\displaystyle\frac{bd}{2\Delta x}\right)\left(\theta e^{0}_{i+1}+\left(1-\theta\right)e^{1}_{i+1}\right)+\left(\displaystyle\frac{2ad}{\Delta x^{2}}+cd\right)\left(\theta e^{0}_{i}+\left(1-\theta\right)e^{1}_{i}\right)- \\ \left(\displaystyle\frac{ad}{\Delta x^{2}}-\displaystyle\frac{bd}{2\Delta x}\right)\left(\theta e^{0}_{i-1}+\left(1-\theta\right)e^{1}_{i-1}\right)= e_{i}^{0}-e_{i}^{1},\\
-\left(\displaystyle\frac{ad}{\Delta x^{2}}+\displaystyle\frac{bd}{2\Delta x}\right)\left(\theta e^{k}_{i+1}+\left(1-\theta\right)e^{k+1}_{i+1}\right)+\left(\displaystyle\frac{2ad}{\Delta x^{2}}+cd\right)\left(\theta e^{k}_{i}+\left(1-\theta\right)e^{k+1}_{i}\right)- \\ \left(\displaystyle\frac{ad}{\Delta x^{2}}-\displaystyle\frac{bd}{2\Delta x}\right)\left(\theta e^{k}_{i-1}+\left(1-\theta\right)e^{k+1}_{i-1}\right)=b_{k}e_{i}^{0}-e_{i}^{k+1}+\displaystyle\sum_{j=0}^{k-1}\left(b_{j}-b_{j+1}\right)e_{i}^{k-j},\\
e_{0}^{k}=e_{n}^{k}=0,
\end{cases}
\end{equation}
where $k\geq1$, $i=1,2,\dots ,n-1$.
We introduce the following grid function:\\
$\displaystyle e^{k}(x)=$
\begin{equation*}
\begin{cases}
\displaystyle e_{l}^{k}\text{, }x\in\left(x_{l-1/2},x_{l+1/2}\right],\text{ }l=1,2\dots ,n-1,\\
\displaystyle 0\text{, }x\in\left(x_{min},x_{min}+\Delta x/2\right]\cup\left[x_{max}-\Delta x/2,x_{max}\right].
\end{cases}
\end{equation*}
Because $\displaystyle e^{k}_{0}=e^{k}_{n}$, we make a periodic expansion for $e^{k}_{l}$ with period $Y=x_{max}-x_{min}$. 
Then $\displaystyle e^{k}(x)$ has the following Fourier series extension:
\begin{equation*}
\displaystyle e^{k}(x)=\displaystyle\sum_{j=-\infty}^{\infty}v_{j}^{k}\exp({2j\pi x i/Y}),
\end{equation*}
where $\displaystyle v_{j}^{k}=\frac{1}{Y}\int_{0}^{Y}e^{k}(x)\exp({2j\pi x i/Y})dx$, $i=\sqrt{-1}$, $k=0, 1,\dots N$.
We define the norm $\norm{\cdot}_{\Delta x}$ as \[\norm{e^{k}}_{\Delta x}=\sqrt{\sum_{j=1}^{n-1}\Delta x\left|e_{j}^{k}\right|^{2}},\] where 
\[e^{k}=(e_{1}^{k},e^{k}_{2},\dots ,e_{n-1}^{k}). \]
Because $e_{0}^{k}=e_{n}^{k}=0,$ it follows
\[\norm{e^{k}}^{2}_{\Delta x}=\int_{0}^{Y}\left|e^{k}(x)\right|^{2}dx=\norm{e^{k}(x)}^{2},\]
where $\norm{\cdot}$ is $L^{2}[0,Y].$\\
Using the Parseval identity we have:
\[\norm{e^{k}}^{2}_{\Delta x}=\sum_{j=1}^{n-1}\Delta x\left|e_{j}^{k}\right|^{2}=Y\sum_{j=-\infty}^{\infty}\left|v_{j}^{k}\right|^{2},\]
$k=0,1,\dots ,N.$ Based on the above analysis and the fact that $x_{l}=x_{min}+lh$, we 
infer that the solution of $(13)$,
has the form:
\begin{equation}
 \displaystyle e_{l}^{k}=v^{k}e^{i\lambda (x_{min}+lh)}, 
\end{equation}
where $\displaystyle \lambda=\frac{2\pi l}{Y}.$
Substituting into $(13)$ we get:
\begin{equation}
\begin{cases}
\displaystyle \left(-\left(\frac{ad}{\Delta x^{2}}+\frac{bd}{2\Delta x}\right)e^{i\lambda \Delta x}+\left(2\frac{ad}{\Delta x^{2}}+cd\right)-\left(\frac{ad}{\Delta x^{2}}-\frac{bd}{2\Delta x}\right)
e^{-i\lambda \Delta x}\right)\left(\theta v^{0}+\left(1-\theta\right)v^{1}\right)=v^{0}-v^{1},\\
\displaystyle \left(-\left(\frac{ad}{\Delta x^{2}}
+\frac{bd}{2\Delta x}\right)e^{i\lambda \Delta x}+
\left(2\frac{ad}{\Delta x^{2}}+cd\right)
-\left(\frac{ad}{\Delta x^{2}}-\frac{bd}{2\Delta x}\right)
e^{-i\lambda \Delta x}\right)
\displaystyle\left(\theta v^{k}+\left(1-\theta\right)v^{k+1}\right)=
\displaystyle\sum_{n=0}^{k-1}\displaystyle\left(b_{n}-b_{n+1}\right)v^{k-n}+b_{k}v^{0}-v^{k+1},\text{ }k\geq 1.
\end{cases}
\end{equation}
To continue we have to find a relation between coefficients $b_{j}$.
\begin{Proposition}
 Coefficients $b_{j}=(j+1)^{1-\alpha}-j^{1-\alpha}$ satisfy:
\begin{enumerate}
 \item $b_{j}>0,$ $j=0,1\dots $
 \item $1=b_{0}>b_{1}>\dots >b_{k}$
 \item $\displaystyle\lim_{k\to\infty}b_{k}=0$
 \item $\displaystyle\sum_{j=0}^{k-1}(b_{j}-b_{j+1})+b_{k}=1$
\end{enumerate}
\end{Proposition} 
\begin{Proof}
 \begin{enumerate}
 \item $b_{j}=(j+1)^{1-\alpha}-j^{1-\alpha}>0,$ for $j\geq0$ and $\alpha\in(0,1)$. 
 \item For $x\geq0$ let us take consider the function $b(x)=(x+1)^{1-\alpha}-x^{1-\alpha}$. Note that $b^{'}(x)=(1-\alpha)((x+1)^{-\alpha}-x^{-\alpha})<0$, so the function is strictly decreasing for $x\geq0$.
 \item It is the consequence of $(1)$ and $(2)$ because strictly decreasing sequence of positive coefficients is converging to $0$.
 \item $\displaystyle\sum_{j=0}^{k-1}(b_{j}-b_{j+1})+b_{k}=(1-b_{1})+(b_{1}-b_{2})+(b_{2}-b_{3})+\dots+(b_{k-1}-b_{k})+b_{k}=1.$
\end{enumerate}
\end{Proof}

Now we will check under which conditions $\left|v^{n}\right|\leq\left|v^{0}\right|$ for each $n=1,\dots ,N$. Then $\norm{e^{k}}\leq \norm{e^{0}}$,
in other words the weighted scheme is stable.
\begin{Theorem}
 Let $\theta\in[0,1)$. If
\begin{enumerate}[label=(\roman*)]
\item \[ 1-\log_{2}\left(2-\frac{\theta}{1-\theta}\right)\leq\alpha,\] or
\item  $\displaystyle 1-\log_{2}\left(2-\frac{\theta}{1-\theta}\right)>\alpha$ or $\theta=1$, and the inequality
 \end{enumerate}
 \begin{equation}
    \displaystyle d\left(\theta-\left(1-\theta\right)\left(b_{0}-b_{1}\right)\right)\left(\left(\frac{4a}{\Delta x^{2}}+c\right)^{2}+\left(\frac{b}{\Delta x}\right)^{2}\right)\leq 2c\left(b_{0}-b_{1}\right),
   \end{equation}
 holds, then the scheme $(11)$ is stable.

 \end{Theorem}
 
 \begin{Proof}
We have to show that $v^{n}$ defined in $(14)$ follows
 $\displaystyle\left|v^{n}\right|\leq\left|v^{0}\right|$ for $n=1,2,\dots ,k.$ 
Let us denote\\
\begin{multline*}
\displaystyle\zeta=\left(-4\sin^{2}\left(\frac{\lambda \Delta x}{2}\right)+2\right)\left(\frac{-ad}{\Delta x^{2}}\right)+2\frac{ad}{\Delta x^{2}}+cd-2i\frac{bd}{2\Delta x}\sin\left(\lambda \Delta x\right)=
 \sin^{2}\left(\frac{\lambda \Delta x}{2}\right)\frac{4ad}{\Delta x^{2}}+cd-i\frac{bd}{\Delta x}\sin\left(\lambda \Delta x\right).
 \end{multline*}
 Let us observe that $\displaystyle Re{\text{ }\zeta}=\sin^{2}\left(\frac{\lambda \Delta x}{2}\right)\frac{4ad}{\Delta x^{2}}+cd>0$. The proof of this fact is immediate because $a,d,c,\Delta x>0$.
 
At the beginning we will show that both statements imply
  \begin{equation}
 \displaystyle\left|\frac{b_{0}-b_{1}-\zeta\theta}{\zeta\left(1-\theta\right)+1}\right|\leq b_{0}-b_{1}.
\end{equation}
Let us assume the first statement. Then $\displaystyle 1-\log_{2}\left(2-\frac{\theta}{1-\theta}\right)\leq\alpha$
is equivalent to $\displaystyle \theta-\left(1-\theta\right)\left(b_{0}-b_{1}\right)\leq0.$ 
\begin{multline*}
\left(\theta^{2}-\left(1-\theta\right)^{2}\left(b_{0}-b_{1}\right)^{2}\right)\left(\left(\sin^{2}\left(\frac{\lambda \Delta x}{2}\right)\frac{4ad}{\Delta x^{2}}+cd\right)^{2}+\left(\sin\left(\lambda \Delta x\right)\frac{bd}{\Delta x}\right)^{2}\right)\\-2\left(b_{0}-b_{1}\right)\left(\left(\sin^{2}\left(\frac{\lambda \Delta x}{2}\right)\frac{4ad}{\Delta x^{2}}+cd\right)\left(\theta+\left(1-\theta\right)\left(b_{0}-b_{1}\right)\right)\right)\leq 0
\end{multline*}

So \[2\left(b_{0}-b_{1}\right)cd-\left(\theta-\left(1-\theta\right)\left(b_{0}-b_{1}\right)\right)\left(cd\right)^{2}\geq0\]
holds for each $\Delta t$, $\Delta x>0.$ Let us observe that\begin{multline*}
\displaystyle0\leq2\left(b_{0}-b_{1}\right)cd-\left(\theta-\left(1-\theta\right)\left(b_{0}-b_{1}\right)\right)\left(cd\right)^{2}\leq\\
2\left(b_{0}-b_{1}\right)\left(\sin^{2}\left(\frac{\lambda \Delta x}{2}\right)\frac{4ad}{\Delta x^{2}}+cd\right)-
\left(\theta-\left(1-\theta\right)\left(b_{0}-b_{1}\right)\right)\left(\sin^{2}\left(\frac{\lambda \Delta x}{2}\right)\frac{4ad}{\Delta x^{2}}+cd\right)^{2}.\end{multline*}
It follows that:
\begin{multline*}
0\leq 2\left(b_{0}-b_{1}\right)\left(\theta+\left(1-\theta\right)\left(b_{0}-b_{1}\right)\right)\left(\sin^{2}\left(\frac{\lambda \Delta x}{2}\right)\frac{4ad}{\Delta x^{2}}+cd\right)-
\left(\theta^{2}-\left(1-\theta\right)^{2}\left(b_{0}-b_{1}\right)^{2}\right)\left(\sin^{2}\left(\frac{\lambda \Delta x}{2}\right)\frac{4ad}{\Delta x^{2}}+cd\right)^{2}\leq\\
2\left(b_{0}-b_{1}\right)\left(\theta+\left(1-\theta\right)\left(b_{0}-b_{1}\right)\right)\left(\sin^{2}\left(\frac{\lambda \Delta x}{2}\right)\frac{4ad}{\Delta x^{2}}+cd\right)-\left(\theta^{2}-\left(1-\theta\right)^{2}\left(b_{0}-b_{1}\right)^{2}\right)\left(\left(\sin^{2}\left(\frac{\lambda \Delta x}{2}\right)\frac{4ad}{\Delta x^{2}}+cd\right)^{2}+\left(\sin\left(\lambda \Delta x\right)\frac{bd}{\Delta x}\right)^{2}\right).\end{multline*}

Note that the right-hand side expression higher than $0$ is equivalent to $(17)$.
Let us assume the second statement.
 The $(16)$ is equivalent to \[\displaystyle 0\leq2cd\left(b_{0}-b_{1}\right)-
\left(\theta-\left(1-\theta\right)\left(b_{0}-b_{1}\right)\right)\left(\left( \frac{4ad}{\Delta x^{2}}+cd\right)^{2}+\left(\frac{bd}{\Delta x}\right)^{2}\right).\]
 Let us observe that if $\displaystyle 1-\log_{2}\left(2-\frac{\theta}{1-\theta}\right)>
 \alpha$ or $\theta=1$, then \[\theta-\left(1-\theta\right)\left(b_{0}-b_{1}\right)>0.\] So
 \begin{multline*}\displaystyle 0\leq2cd\left(b_{0}-b_{1}\right)-\left(\theta-\left(1-\theta\right)\left(b_{0}-b_{1}\right)\right)\left(\left(\frac{4ad}{\Delta x^{2}}+cd\right)^{2}+\left(\frac{bd}{\Delta x}\right)^{2}\right)\leq\\
 2cd\left(b_{0}-b_{1}\right)-
\left(\theta-\left(1-\theta\right)\left(b_{0}-b_{1}\right)\right)\left(\left( \sin^{2}\left(\frac{\lambda \Delta x}{2}\right)\frac{4ad}{\Delta x^{2}}+cd\right)^{2}+\left(\frac{bd}{\Delta x}\sin \lambda \Delta x\right)^{2}\right).
\end{multline*}
Similarly as it was done for the first statement, it can be shown that the right-hand side expression higher than $0$ implies $(17)$.
 
We will follow the mathematical induction method to show that for each $n=1,2\dots,N$ there holds $\left|v^{n}\right|\leq\left|v^{0}\right|$.
 \begin{enumerate}
 \item $n=1$
 By the identity \[ \sin^{2}\frac{z}{2}=-\frac{1}{4}\left(e^{iz}-2+e^{-iz}\right),\]
 the first equation of $(15)$ has the form\\
 \begin{equation*}\left(\left(-4\sin^{2}\left(\frac{\lambda \Delta x}{2}\right)+2\right)\left(\frac{-ad}{\Delta x^{2}}\right)+\left(2\frac{ad}{\Delta x^{2}}+cd\right)-2i\frac{bd}{2\Delta x}\sin\left(\lambda \Delta x\right)\right)\cdot \displaystyle\left(\left(1-\theta\right) v^{1}+\theta v^{0}\right)=
 v^{0}-v^{1}.\end{equation*}
It is equivalent to
 \[\zeta\left(\left(1-\theta\right)v^{1}+\theta v^{0}\right)=v^{0}-v^{1},\]
  \[\left(\zeta\left(1-\theta\right)+1\right)v^{1}=\left(1- \zeta \theta\right)v^{0}.\]
  So 
   \[\displaystyle\left|v^{1}\right|=\left|\frac{1-\zeta \theta}{\zeta\left(1-\theta\right)+1}\right| \left|v^{0}\right|.\]
   It is easy to check that $(17)$ implies $\displaystyle\left|\frac{1-\zeta \theta}{\zeta\left(1-\theta\right)+1}\right|\leq1$, so
   \[\displaystyle\left|v^{1}\right|\leq \left|v^{0}\right|.\]
  \item Let us suppose that
  \[\displaystyle\left|v^{n}\right|\leq\left|v^{0}\right|,\]
  for $n=1,2,\dots ,k,$ $k<N.$\\
  To complete the proof we have to show that 
  \[\displaystyle\left|v^{k+1}\right|\leq\left|v^{0}\right|.\]
  By the second equation of $(15)$, for $k\geq1$ we have:
  \[\displaystyle \zeta\left(\left(1-\theta\right) v^{k+1}+\theta v^{k}\right)=-v^{k+1}+\sum_{n=0}^{k-1}\left(b_{n}-b_{n+1}\right)v^{k-n}+b_{k}v^{0}, \]
  it is equivalent to 
  \[v^{k+1}\left(\left(1-\theta\right) \zeta+1\right)=-\theta \zeta v^{k}+\sum_{n=0}^{k-1}\left(b_{n}-b_{n+1}\right)v^{k-n}+b_{k}v^{0}.\]
  So \[\displaystyle \left|v^{k+1}\right|\left|\left(\left(1-\theta\right) \zeta+1\right)\right|\leq 
  \left|\left(b_{0}-b_{1}-\theta\zeta\right)\right| \left|v^{k}\right|
  +\sum_{n=1}^{k-1}\left(b_{n}-b_{n+1}\right)\left|v^{k-n}\right|+b_{k}\left|v^{0}\right|.
\]

 Dividing by $\left|\left(1-\theta\right)\zeta+1\right|$ we get
 \begin{multline*}
 \displaystyle \left|v^{k+1}\right|\leq 
 \left|\frac{\left(b_{0}-b_{1}-\theta\zeta\right)}{\left(1-\theta\right) \zeta+1}\right| \left|v^{k}\right|+
 \left|\frac{\sum\limits_{n=1}^{k-1}\left(b_{n}-b_{n+1}\right)}{\left(1-\theta\right) \zeta+1}\right|\left|v^{k-n}\right|+
 \frac{b_{k}}{\left|\left(1-\theta\right) \zeta+1\right|}\left|v^{0}\right|\leq\\
  \left|\frac{\left(b_{0}-b_{1}-\theta\zeta\right)}{\left(1-\theta\right) \zeta+1}\right| \left|v^{0}\right|+
 \left|\sum_{n=1}^{k-1}\left(b_{n}-b_{n+1}\right)\right|\left|v^{0}\right|+
 b_{k}\left|v^{0}\right|\leq \left(b_{0}-b_{1}\right) \left|v^{0}\right|+ \left(\sum_{n=1}^{k-1}\left(b_{n}-b_{n+1}\right)+b_{k}\right)\left|v^{0}\right|=\left|v^{0}\right|
, 
\end{multline*}
where the second inequality holds because $\displaystyle Re{\text{ }\zeta}>0$ and the latest by $(17)$. As a result we have
    \[\left|v^{k+1}\right|\leq\left|v^{0}\right|.\]
  \end{enumerate}
  By the mathematical induction method the proof is completed.
\end{Proof}\\ 
In particular, the implicit scheme is unconditionally stable for each $\alpha\in\left(0,1\right)$. Similarly, the explicit and the C-N schemes are conditionally stable for each $\alpha\in\left(0,1\right)$.

\subsection{Convergence of the weighted discrete scheme}
Let us denote
$u_{l}^{k}=u(x_{l},t_{k})$ - the exact solution of $(6)$ evaluated at the grid point,
$\hat{u}_{l}^{k}$ - the solution of the numerical scheme $(11)$.
Let us define the error at the point $(x_{l},t_{k})$ by
$E_{l}^{k}=u_{l}^{k}-\hat{u}_{l}^{k}$, $l=0,1,\dots n$, $k=0,1,..,N$. \\
Similarly, as $(13)$ we get the following system:
\begin{equation}
\begin{cases}
-\left(\displaystyle\frac{ad}{\Delta x^{2}}+\displaystyle\frac{bd}{2\Delta x}\right)\left(\theta E^{0}_{i+1}+\left(1-\theta\right)E^{1}_{i+1}\right)+\left(\displaystyle\frac{2ad}{\Delta x^{2}}+cd\right)\left(\theta E^{0}_{i}+\left(1-\theta\right)E^{1}_{i}\right)\\-
\left(\displaystyle\frac{ad}{\Delta x^{2}}-\displaystyle\frac{bd}{2\Delta x}\right)\left(\theta E^{0}_{i-1}+\left(1-\theta\right)E^{1}_{i-1}\right)=
E_{i}^{0}-E_{i}^{1}+\Delta t^{\alpha}\left(\theta R_{i}^{0}+\left(1-\theta\right)R_{i}^{1}\right),\\
-\left(\displaystyle\frac{ad}{\Delta x^{2}}+\displaystyle\frac{bd}{2\Delta x}\right)\left(\theta E^{k}_{i+1}+\left(1-\theta\right)E^{k+1}_{i+1}\right)+\left(\displaystyle\frac{2ad}{\Delta x^{2}}+cd\right)\left(\theta E^{k}_{i}+\left(1-\theta\right)E^{k+1}_{i}\right)\\- \left(\displaystyle\frac{ad}{\Delta x^{2}}-\displaystyle\frac{bd}{2\Delta x}\right)\left(\theta E^{k}_{i-1}+\left(1-\theta\right)E^{k+1}_{i-1}\right)=b_{k}E_{i}^{0}-E_{i}^{k+1}+\displaystyle\sum_{j=0}^{k-1}\left(b_{j}-b_{j+1}\right)E_{i}^{k-j}+\\
 \displaystyle\Delta t^{\alpha}\left(\theta R_{i}^{k}+\left(1-\theta\right)R_{i}^{k+1}\right),\\
E_{i}^{0}=0,\\
E_{0}^{k}=E_{n}^{k}=0,
\end{cases}
\end{equation}
where $k=1,\ldots,N$, $i=1,2,\dots ,n-1$ and $\displaystyle\theta\in\left[0,1\right]$.
Similarly as in case of the stability we will proceed with the von Neumann method.

We introduce the following grid functions:\\
$E^{k}(x)=$
\begin{equation*}
\begin{cases}
\displaystyle E_{l}^{k}\text{, }x\in\left(x_{l-1/2},x_{l+1/2}\right],\text{ }l=1,2\dots ,n-1,\\
\displaystyle 0\text{, }x\in\left(x_{min},x_{min}+\Delta x/2\right]\cup\left[x_{max}-\Delta x/2,x_{max}\right].
\end{cases}
\end{equation*}

$R^{k}(x)=$
\begin{equation*}
\begin{cases}
\displaystyle R_{l}^{k}\text{, }x\in\left(x_{l-1/2},x_{l+1/2}\right],\text{ }l=1,2\dots ,n-1,\\
\displaystyle 0\text{, }x\in\left(x_{min},x_{min}+\Delta x/2\right]\cup\left[x_{max}-\Delta x/2,x_{max}\right].
\end{cases}
\end{equation*}
Because $E^{k}_{0}=E^{k}_{n}$, we make a periodic expansion for $E^{k}_{l}$ with the period $Y=x_{max}-x_{min}$. 
Then $E^{k}(x)$ has the following Fourier series extension:
\begin{equation*}
\displaystyle E^{k}(x)=\sum_{j=-\infty}^{\infty}w_{j}^{k}\exp({2j\pi x i/Y}),
\end{equation*}
where $\displaystyle w_{j}^{k}=\frac{1}{Y}\int_{0}^{Y}E^{k}(x)\exp({2j\pi x i/Y})dx$, $i=\sqrt{-1}$, $k=0,1,\dots N$.\\
By analogy, because $R^{k}_{0}=R^{k}_{n}$, we make a periodic expansion for $R^{k}_{l}$ with the period $Y$. 
Then $R^{k}(x)$ has the following Fourier series extension:
\begin{equation*}
\displaystyle R^{k}(x)=\sum_{j=-\infty}^{\infty}r_{j}^{k}\exp({2j\pi x i/Y}),
\end{equation*}
where $\displaystyle r_{j}^{k}=\frac{1}{Y}\int_{0}^{Y}R^{k}(x)\exp({2j\pi x i/Y})dx$, $i=\sqrt{-1}$, $k=0,1,\dots N$.

We define the norm $\norm{\cdot}_{\Delta x}$ as \[\displaystyle \norm{E^{k}}_{\Delta x}=\sqrt{\sum_{j=1}^{n-1}\Delta x\left|E_{j}^{k}\right|^{2}},\] \[\norm{R^{k}}_{\Delta x}=\sqrt{\sum_{j=1}^{n-1}\Delta x\left|R_{j}^{k}\right|^{2}},\] where 
\[\displaystyle E^{k}=\left(E_{1}^{k},E^{k}_{2},\dots ,E_{n-1}^{k}\right), \]
\[\displaystyle R^{k}=\left(R_{1}^{k},R^{k}_{2},\dots ,R_{n-1}^{k}\right). \]
Because $E_{0}^{k}=E_{n}^{k}=0,$ and $R_{0}^{k}=R_{n}^{k}=0,$ there holds

\[\displaystyle \norm{E^{k}}_{\Delta x}^{2}=\int_{0}^{Y}\left|E^{k}\left(x\right)\right|^{2}dx=\norm{E^{k}\left(x\right)}^{2}_{L^{2}},\]
\[\displaystyle \norm{R^{k}}_{\Delta x}^{2}=\int_{0}^{Y}\left|R^{k}\left(x\right)\right|^{2}dx=\norm{R^{k}\left(x\right)}^{2}_{L^{2}}.\]
Using the Parseval identity we have:
\begin{equation}
\begin{cases}
 \norm{E^{k}}_{\Delta x}^{2}=\displaystyle\sum_{j=1}^{n-1}\Delta x\left|E_{j}^{k}\right|^{2}=Y\sum_{j=-\infty}^{\infty}\left|w_{j}^{k}\right|^{2},\\
 \norm{R^{k}}_{\Delta x}^{2}=\displaystyle\sum_{j=1}^{n-1}\Delta x\left|R_{j}^{k}\right|^{2}=Y\sum_{j=-\infty}^{\infty}\left|r_{j}^{k}\right|^{2},
 \end{cases}
\end{equation}
where $k=0,1,\dots ,N.$ Based on the above analysis and the fact that $x_{l}=x_{min}+lh$, we 
suppose that the solution of $\left(18\right)$,
has the form:
\[E_{l}^{k}=w^{k}e^{i\lambda \left(x_{min}+lh\right)},\]
\[R_{l}^{k}=r^{k}e^{i\lambda \left(x_{min}+lh\right)},\]
where $\displaystyle\lambda=\frac{2\pi l}{Y}.$
Substituting into $\left(18\right)$ we get:

\begin{equation}
\begin{cases}
\begin{gathered}
\displaystyle \left(-\left(\frac{ad}{\Delta x^{2}}+\frac{bd}{2\Delta x}\right)
e^{i\lambda \Delta x}+\left(2\frac{ad}{\Delta x^{2}}+cd\right)-\left(\frac{ad}{\Delta x^{2}}-\frac{bd}{2\Delta x}\right)
e^{-i\lambda \Delta x}\right)\left(\theta w^{0}+\left(1-\theta\right)w^{1}\right)=\\
\displaystyle w^{0}-w^{1}+\Delta t^{\alpha}\left(\theta r^{0}+\left(1-\theta\right)r^{1}\right),\\
\displaystyle\left(-\left(\frac{ad}{\Delta x^{2}}
+\frac{bd}{2\Delta x}\right)e^{i\lambda \Delta x}+
\left(2\frac{ad}{\Delta x^{2}}+cd\right)
-\left(\frac{ad}{\Delta x^{2}}-\frac{bd}{2\Delta x}\right)
e^{-i\lambda \Delta x}\right)
\left(\theta w^{k}+\left(1-\theta\right)w^{k+1}\right)=\\
\displaystyle\sum_{n=0}^{k-1}\left(b_{n}-b_{n+1}\right)w^{k-n}+b_{k}w^{0}-w^{k+1}+\Delta t^{\alpha}\left(\theta r^{k}+\left(1-\theta\right)r^{k+1}\right),
\end{gathered}
\end{cases}
\end{equation}
where $k=1,\ldots,N-1$. Let us denote \[ \displaystyle\sin^{2}\frac{z}{2}=-\frac{1}{4}\left(e^{iz}-2+e^{-iz}\right).\] 
Then, taking into account that $r^{0}=0$ and $w^{0}=0$, $(20)$ has the form

\begin{equation}
\begin{cases}
\displaystyle \zeta\left(1-\theta\right)w^{1}=-w^{1}
 +\Delta t^{\alpha}\left(1-\theta\right)r^{1},\\
  \displaystyle\zeta\left(\left(1-\theta\right) w^{k+1}+\theta w^{k}\right)=-w^{k+1}+
  \displaystyle\sum_{n=0}^{k-1}\left(b_{n}-b_{n+1}\right)w^{k-n}+\Delta t^{\alpha}\left(\theta r^{k}+\left(1-\theta\right)r^{k+1}\right),
  \end{cases}
\end{equation}
where $k=1,\ldots,N-1$, $\theta\in[0,1]$ and $\zeta$ is previously defined.

\begin{Lemma}
 If condition $(i)$ or $(ii)$ in Theorem $3.2$ is satisfied, then 
$w^{k}$ follows 
 \[\displaystyle\left|w^{k+1}\right|\leq \frac{\max\left(1-\theta,C_{1}\right)}{b_{k}}\Delta t^{\alpha}\left|r^{1}\right|.\]

where $k=0,1,\dots ,N-1$ and the constant $C_{1}$ is independent of  $\theta$, $\Delta t$ and $\Delta x$.

\end{Lemma}

\begin{Proof}
 Because $\displaystyle R_{l}^{k}=O\left(\Delta t^{2-\alpha}+\Delta x^{2}\right),$ so there exists a positive constant $C_{1}$, such that
 \[\displaystyle\left|R_{l}^{k}\right|\leq C_{1}\left(\Delta t^{2-\alpha}+\Delta x^{2}\right).\]
 Then 
 \begin{equation}
  \displaystyle\norm{R^{k}}_{\Delta x}\leq C_{1}\sqrt{Y}\left(\Delta t^{2-\alpha}+\Delta x^{2}\right).
 \end{equation}

 Convergence of the right series in second line of $\left(19\right)$ implies that
\[\displaystyle\left|r^{k}\right|=\left|R_{l}^{k}\right|\leq C_{1}\left|R_{l}^{1}\right|=C_{1}\left|r^{1}\right|.\]
 
 Then by $(21)$ and Proposition 3.1 
 we have \[\displaystyle \left|w^{1}\right|= \frac{\left(1-\theta\right)}{\left|\zeta\left(1-\theta\right)+1\right|}\left|\Delta t^{\alpha}r^{1}\right|\leq \frac{\left(1-\theta\right)}{b_{0}}\Delta t^{\alpha}\left|r^{1}\right| \leq \frac{\max\left(1-\theta,C_{1}\right)}{b_{0}}\Delta t^{\alpha}\left|r^{1}\right|.\]
 The first inequality is true because $Re{\text{ }\zeta}>0$.
 Now let us suppose that    
 \[\displaystyle \left|w^{n}\right|\leq \frac{\max\left(1-\theta,C_{1}\right)}{b_{k-1}}\Delta t^{\alpha}\left|r^{1}\right|,\]
 where $n=2,3,\dots ,k$ and $C_{1}$ is a
 constant independent of $\theta$, $\Delta t$ and $\Delta x$. 
 
 By $(21)$ we have \begin{multline*}
 \displaystyle\left|\left(1-\theta\right)\zeta+1\right| \left|w^{k+1}\right|=\left|-\zeta\theta w^{k}+\sum_{j=0}^{k-1}\left(b_{j}-b_{j+1}\right)w^{k-j}+
 \Delta t^{\alpha} \left(\left(1-\theta\right)r^{k+1}+\theta r^{k}\right)\right|\leq \\\left|1-b_{1}-\zeta\theta\right| \left|w^{k}\right|+\sum_{j=1}^{k-1}\left(b_{j}-b_{j+1}\right)\left|w^{k-j}\right|
 +C_{1}\Delta t^{\alpha} \left|r^{1}\right| \leq\\ \left(\left|1-b_{1}-\zeta\theta\right|\frac{\max\left(1-\theta,C_{1}\right)}{b_{k-1}}+\sum_{j=1}^{k-1}\left(b_{j}-b_{j+1}\right)\frac{\max\left(1-\theta,C_{1}\right)}{b_{k-1}}
+\max\left(1-\theta,C_{1}\right)\right)\Delta t^{\alpha}\left|r^{1}\right|\leq\\  \left(\left|1-b_{1}-\zeta\theta\right|+\sum_{j=1}^{k-1}\left(b_{j}-b_{j+1}\right)+b_{k}\right)\frac{\max\left(1-\theta,C_{1}\right)}{b_{k}}\Delta t^{\alpha}\left|r^{1}\right|,
\end{multline*} 
 \\Dividing by the coefficient $\left|\left(1-\theta\right)\zeta+1\right|$ , we get
\begin{multline*}
 \displaystyle\left|w^{k+1}\right|\leq \left(\left|\frac{b_{0}-b_{1}-\theta \zeta}{\zeta\left(1-\theta\right)+1}\right|+\frac{\displaystyle\sum\limits_{j=1}^{k-1}\left(b_{j}-b_{j+1}\right)+b_{k}}{\left|\zeta\left(1-\theta\right)+1\right|}\right)\frac{\max\left(1-\theta,C_{1}\right)}{b_{k}}\Delta t^{\alpha}\left|r^{1}\right|\leq
 \\\left(b_{0}-b_{1}+\sum_{j=1}^{k-1}\left(b_{j}-b_{j+1}\right)+b_{k}\right)\frac{\max\left(1-\theta,C_{1}\right)}{b_{k}}\Delta t^{\alpha}\left|r^{1}\right|=\frac{\max\left(1-\theta,C_{1}\right)}{b_{k}}\Delta t^{\alpha}\left|r^{1}\right|.
\end{multline*}
The last inequality is true by $(17)$ and because $Re{\text{ }\zeta}>0$.

 By the mathematical induction the proof is completed.
\end{Proof}

\begin{Theorem}
 If condition $(i)$ or $(ii)$ in Theorem $3.2$ is satisfied, then the discrete scheme $\left(11\right)$ is convergent and it follows that
 \[\displaystyle\left|u_{l}^{k}-\hat{u}_{l}^{k}\right|\leq \max\left(1-\theta, C_{1}\right)C_{2} \left(\Delta t^{2-\alpha}+\Delta x^{2}\right),\]
 for $k=1,2,\dots ,N,$ where $C_{1}$ and $C_{2}$ are positive constants independent of $\theta$, $\Delta t$ and $\Delta x$.
\end{Theorem}

\begin{Proof}
 Let us observe that $b_{k-1}^{-1}k^{-\alpha}\leq 1/(1-\alpha)$, $k=1,2,\dots ,N$. By the Lemma $3.1$
 \[\displaystyle\left|w^{k}\right|\leq \frac{\max\left(1-\theta, C_{1}\right)}{b_{k-1}}\Delta t ^{\alpha}\left|r^{1}\right|=\frac{\max\left(1-\theta, C_{1}\right)}{b_{k-1}}\Delta t ^{\alpha}k^{\alpha}k^{-\alpha}\left|r^{1}\right|\leq
 \frac{\max\left(1-\theta, C_{1}\right)}{1-\alpha}\left(\Delta t k\right)^{\alpha}\left|r^{1}\right|\leq \frac{\max\left(1-\theta, C_{1}\right)}{1-\alpha}T^{\alpha}\left|r^{1}\right|.\] 
 Similarly, by $\left(19\right)$ and $\left(22\right)$ we have the following:
\[\displaystyle\norm{E^{k}}_{\Delta x}\leq \frac{\max\left(1-\theta, C_{1}\right)}{1-\alpha}T^{\alpha}\norm{R^{1}}_{\Delta x}\leq \frac{\max\left(1-\theta, C_{1}\right)}{1-\alpha}T^{\alpha}\sqrt{Y}\left(\Delta t^{2-\alpha}+\Delta x^{2}\right).\]
After taking $\displaystyle C_{2}=\frac{T^{\alpha}\sqrt{Y}}{1-\alpha}$ the proof is completed.
\end{Proof}

\begin{figure}[h]
\centering
\includegraphics[scale=0.56]{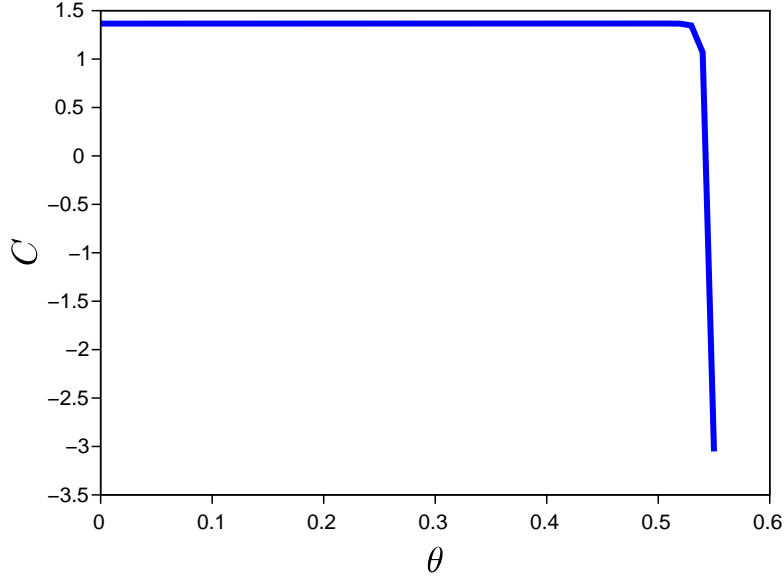}
\caption{The dependence of the European call price on $\theta$. The jump into negative regime is the result 
of the lack of the unconditional stability beyond of the interval $[0,\check{\theta_{\alpha}}]$. The parameters are $n=5000$, $\sigma=1$, $N=140$, $x_{max}=10$, $x_{min}=-20$, $T=4$, $K=1$, $Z_{0}=2$, $r=0.04$, $\alpha=0.5$.}   
\end{figure}\newpage
Let us observe that as a direct conclusion of Theorem 3.3 we get, that the optimal choice of $\theta$ for given $\alpha$ is such that $\displaystyle\log_{2}\left(2-\frac{\check{\theta_{\alpha}}}{1-\check{\theta_{\alpha}}}\right)=1-\alpha,$ equivalently $\displaystyle\check{\theta_{\alpha}}=\frac{2-2^{1-\alpha}}{3-2^{1-\alpha}}$. Then the lowest boundary for an error is achieved without loosing the unconditional stability/convergence. In Figure $2$ the jump into negative values is the result of the increasing error, what the consequence of lack of the stability is. 
Note that in the case of classical B-S model ($\alpha=1$) for implicit scheme, the method has $(\Delta x^{2}+\Delta t)$, but for $\check{\theta_{1}}=1/2$ the method has $(\Delta x^{2}+\Delta t^{2})$ order of convergence \cite{ja}.
Similarly the C-N scheme is unconditionally convergent only for $\alpha=1$. Example 2 confirms that for $\alpha$ close to $1$ the C-N has the lowest numerical error without accelerating time of computation. \\
\\As it is shown in \cite{problem}, in the similar class of problems the solution has a weak singularity near the
initial time $t=0$. Due to this fact, the order of convergence with respect to time can fall from the value of $2-\alpha$ to $\alpha$. In our case the solution $u$ has no singularities in the considered domain. It is clear from the direct interpretation of $u$ as a price of an option, which for initial time $t=0$ is equal to a payoff function.\\
\\ The case $\theta=0$ is the implicit numerical scheme investigated in \cite{zhang}. Our work confirms the previous results and generalize them for other types of numerical schemes. We also 
find the optimal value of parameter $\theta$ for the following class of fractional-type problems
\begin{equation*}
 \begin{cases}
{}_{0}^{c}D_{t}^{\alpha}u\left(x,t\right)= a\displaystyle\frac{\partial^2 u\left(x,t\right)}{\partial x^2}+b\displaystyle\frac{\partial u\left(x,t\right)}{\partial x}+cu\left(x,t\right),\\
u\left(x,0\right)=f\left(x\right),\\
u\left(x_{min},t\right)=p\left(t\right),\\
u\left(x_{max},t\right)=q\left(t\right),\\
\end{cases}
\end{equation*}
for $\left(x,t\right)\in\mathbb{X}\times\left(0,T\right)$. Here, $\mathbb{X}=[x_{min},x_{max}]$ is a fixed interval, $T$ is a time horizon, $a$, $b$, $c\in\mathbb{R}$, $\alpha\in(0,1)$. We assume that $u$, $q$, $p$, $f$ are smooth enough (see \cite{exist} and references therein).

\subsection{Numerical examples}
\begin{Example}
Let us take parameters $T=1$, $Z_{0}=1$, $\sigma=1$, $r=0.04$, $K=2$. Using the fact that 
\begin{equation*}
 p\approx\log_{2}\frac{\tilde{u}(h)-u}{\tilde{u}(h/2)-u},
\end{equation*}
where $p$ is the order of convergence and $\tilde{u}(h)$ is the solution of the numerical scheme $(11)$ (evaluated at the fixed point, similarly as $u$) for the length
of the grid equal to $h$, we can numerically check the order of convergence (with respect to each variable) of the numerical scheme. 
The comparison prepared for both variables represent Table 1 and Table 2. The empirical order related to $\Delta t$ and $\Delta x$ should be close to $2-\alpha$ and $2$ respectively.

\begin{table}[!ht]\footnotesize
\begin{center}
\begin{tabular}{|c c c c c |} 
\hline
$\alpha$& $h_{\Delta t}$&   empirical order  &  theoretical value&relative error\\ \hline
$0.99$& $0,01$ &   $1,02$  &  $1,01$&$1,39\%$\\			
$0.7$& $2,22\times10^{-3}$&   $1,32$  &  $1,3$ &$1,27\%$\\%
$0.5$& $1,67\times10^{-3}$&   $1,51$  &  $1,5$ &$0,67\%$\\%
$0.3$& $1,39\times10^{-3}$&   $1,7$  &  $1,7$ &$0,21\%$\\%
$0.1$& $1,25\times10^{-3}$&   $1,85$  &  $1,9$ &$2,7\%$\\%

\hline
\end{tabular}
\caption{\label{table:1}Order of convergence with respect to $\Delta t$ for $\theta=0$, $\Delta x=0,2$, $x_{max}=1$, $x_{min}=-1$, and different $\alpha$. We approximate $u$ by $\tilde{u}$ calculated for $\Delta t=3,85\times 10^{-4}$. Functions are evaluated at point $x=-0.01$.}
\end{center}
\end{table}

\begin{table}[!ht]\footnotesize
\begin{center}
\begin{tabular}{|c c c c |} 
\hline
$\alpha$&   empirical order  &  theoretical value&relative error\\ \hline
$0.99$ &   $1,97$  &  $2$&$1,64\%$\\			
$0.7$&   $1,96$  &  $2$ &$2,05\%$\\%
$0.5$&   $1,95$  &  $2$ &$2,3\%$\\%
$0.3$&   $1,95$  &  $2$ &$2,52\%$\\%
$0.1$&   $1,95$  &  $2$ &$2,71\%$\\%

\hline
\end{tabular}
\caption{\label{table:2}Order of convergence with respect to $\Delta x$ for $\theta=0$, $\Delta t=0,05$, $x_{max}=10$, $x_{min}=-1$, and different $\alpha$. To show that the order should not depend on $\alpha$, in each case we take $h_{\Delta x}=5,5\times10^{-2}$. We approximate $u$ by $\tilde{u}$ calculated for $\Delta t=3,33\times 10^{-3}$ and $\Delta x=1,83\times 10^{-2}$. Functions are evaluated at point $x=x_{max}$.}
\end{center}
\end{table}
\end{Example}

\begin{Example}
Let us consider the case $\sigma=1$, $r=0.04$, $x_{max}=10$, $x_{min}=-20$, $T=4$, $K=2$, $Z_{0}=1$. The Tables $3$ and $4$ present the comparison of error and time of computation for European call option, in dependence of parameters $\theta$, $n$, $N$. The simulations are made for $\alpha=0.999$ and they are compared to the B-S formula's result equal $0.593$. The most optimal choice of $\theta$ is close to $\check{\theta_{1}}=1/2$. 
\begin{table}[!ht]\footnotesize
\begin{center}
\scalebox{0.94}{
\begin{tabular}{|c c c c c c c|} 
\hline
(n,N)        &(5000,140)&(3000,100)&(500,50)&(100,20)&(200,200)&(50,1300)\\\hline
$\theta=0$&$0.63\%$&$0.87\%$&$1.74\%$&$6.31 \%$&$0.58 \%$&$3.17\%$\\
$\theta=0.25$&$0.43\%$&$0.59\%$&$1.12\%$&$4.87 \%$&$0.44 \%$&$3.15\%$\\
$\theta=0.5$&$0.23\%$&$0.31\%$&$0.61\%$&$2.04\%$&$0.3\%$&$3.13\%$\\
$\theta=0.6$&$6.45\times 10^{6}\%$&$5.23\times10^{4}\%$&$68.34\%$&$2.58 \%$&$0.25 \%$&$3.12\%$\\
$\theta=0.9$&$1.06\times 10^{55}\%$&$6.13\times10^{38}\%$&$2.09\times10^{17}\%$&$85\%$&$1.03\times10^{9}\%$&$3.09\%$\\
\hline
\end{tabular}}\normalsize
\caption{\label{table:3}Finite difference method for different $\theta$, $n$ and $N$.}
\end{center}
\end{table}

\begin{table}[!ht]\footnotesize
\begin{center}
\scalebox{0.94}{
\begin{tabular}{|c c c c c c c|} 
\hline
(n,N)        &(5000,140)&(3000,100)&(500,50)&(100,20)&(200,200)&(50,1300)\\\hline
$\theta=0$&$9.7s$&$3.9s$&$0.7s$&$0.3s$&$3.6s$&$112.4s$\\
$\theta=0.25$&$10.3s$&$4.3s$&$0.7s$&$0.3s$&$3.6s$&$110.6s$\\
$\theta=0.5$&$9.7s$&$3.9s$&$0.6s$&$0.3s$&$3.6s$&$110.4s$\\
$\theta=0.6$&$10.5s$&$4.1s$&$0.8s$&$0.3s$&$3.6s$&$110s$\\
$\theta=0.9$&$10.2s$&$4.1s$&$0.7s$&$0.4s$&$3.5s$&$111s$\\
\hline
\end{tabular}}\normalsize
\caption{\label{table:4}Time related to run Finite difference method for Table 3.}
\end{center}
\end{table}
\end{Example}\newpage

\begin{Example}
 Let us take the parameters $Z_{0}=\sigma=1$, $r=0.04$, $n=1000$, $N=140$, $x_{max}=10$, $x_{min}=-20$, $\theta=\check{\theta_{\alpha}}$. 
 Since $ES_{\alpha}\left(T\right)=\frac{T^{\alpha}}{\Gamma\left(1-\alpha\right)}$, for $T>1$ the price of European call option is increasing function of $\alpha$ (see Figures $4$ and $5$), similarly for $T<1$ the price of European call option is decreasing function of $\alpha$. 
 For $T>1$ it follows our intuition, because with decreasing value of $\alpha$, the constant periods in dynamics of underlying instrument appear more frequently. Such asset can be considered as more predictable, so value of its European call should be lower than the same options on instruments driven by higher values of $\alpha$.
The dependence of the European call option on $T$ and $\alpha$ is presented in Figure $3$. In Figure $5$ we compare the finite difference method (FD) to the Monte Carlo method (MC) introduced in \cite{MM}. MC is oscillating around the FD output. Both methods are efficient and can be used to compute the prices of European options.
 \end{Example}

\begin{figure}[ht]
\centering
\includegraphics[scale=0.52]{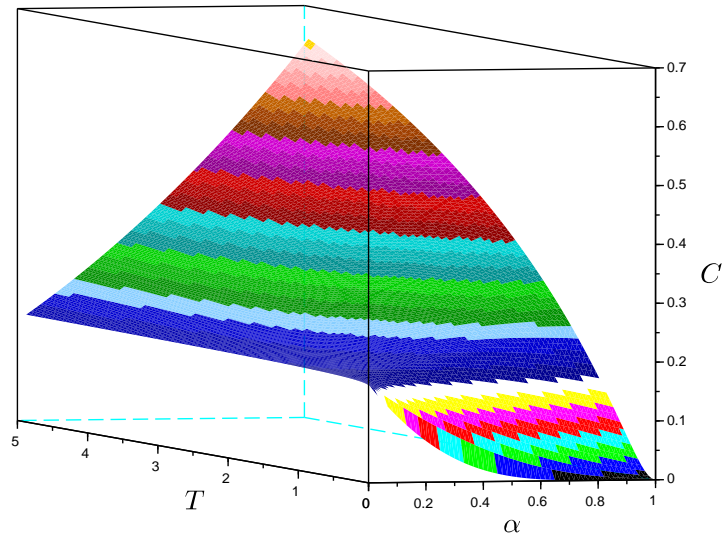}
\caption{The dependence of the European call price on $T$ and $\alpha$. The argument $T=1$ is the inflection point of the function.} 
\end{figure} 
\newpage
\begin{figure}[ht]
\centering

\includegraphics[scale=0.43]{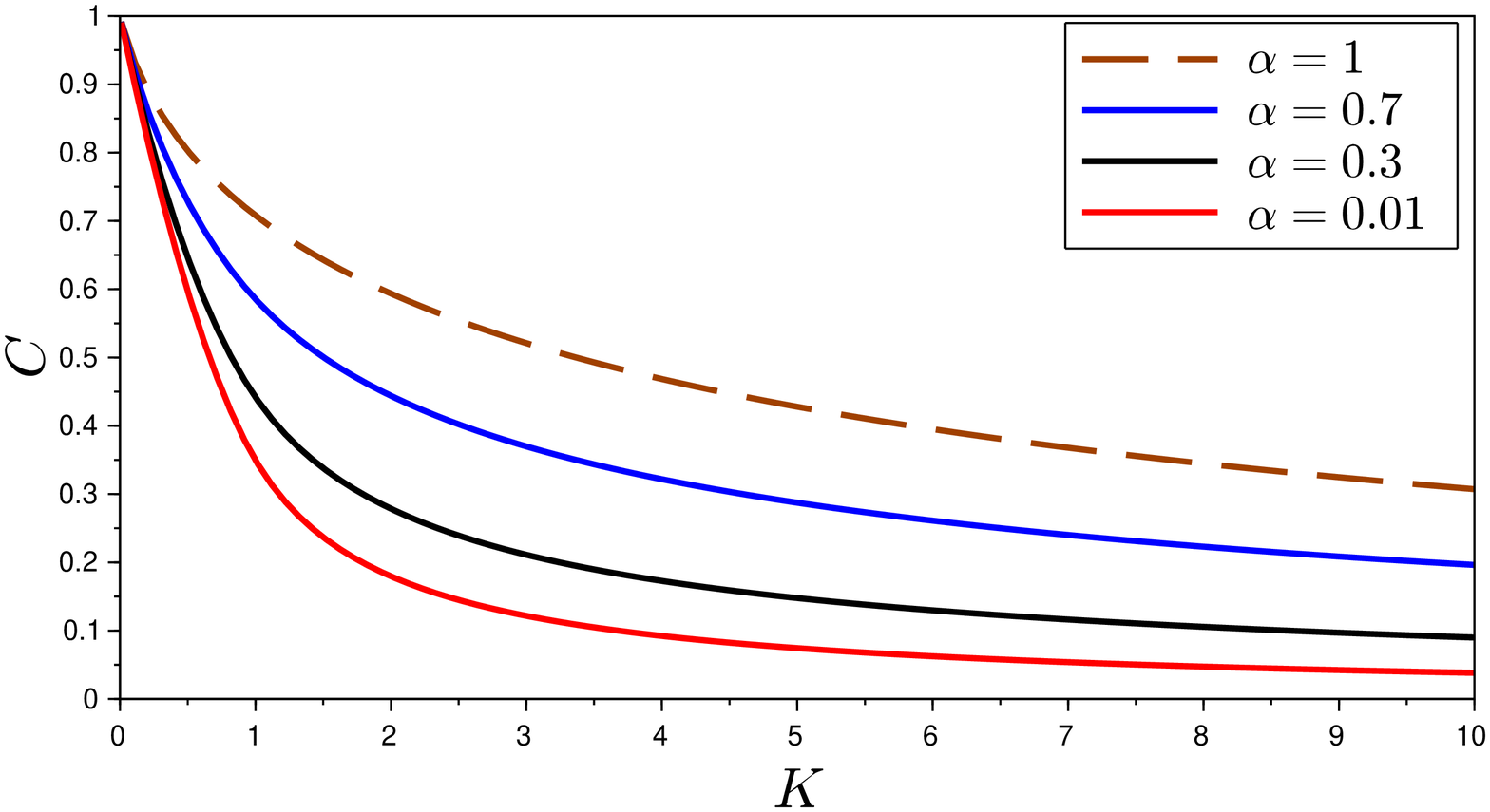}
\caption{The dependence of the European call price on $K$ for $T=4$. }   

 \vspace*{\floatsep}
\includegraphics[scale=0.48]{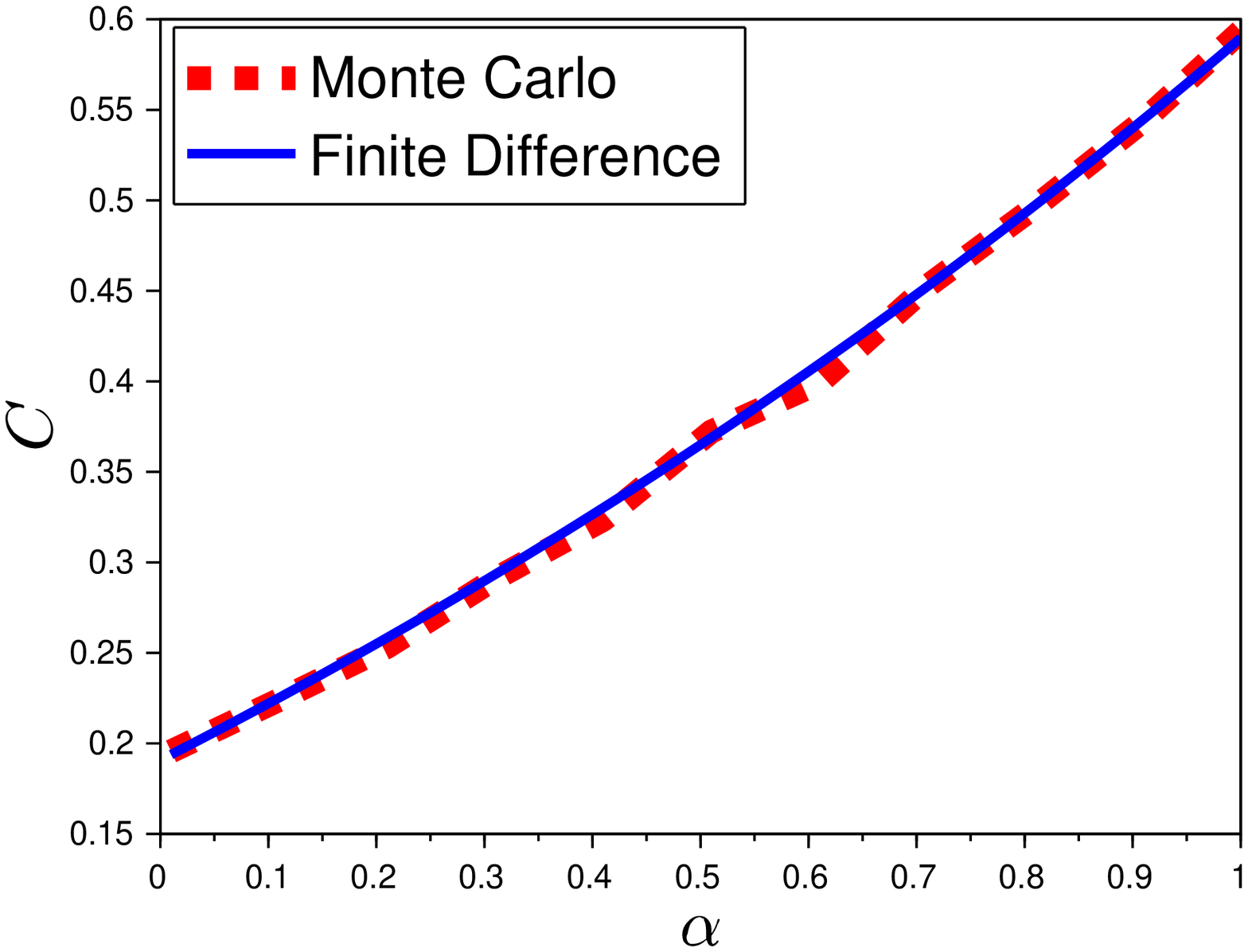}
\caption{The price of European call option in dependence of $\alpha$ for $T=4$. The introduced FD method is compared to the MC method explained in \cite{MM} for $M=800$ repetitions. Increasing $M$ follows the result of MC will approach the FD output. The strike $K=2$.}   
\end{figure} 
\section{Summary}
In this paper:
\begin{itemize}
\item[--] We have shown that the solution of fractional B-S equation is equal to the fair price of European option with respect to $\mathbb{Q}$ in subdiffusive B-S model.
\item[--] We have introduced weighted numerical scheme for this equation. It allows us to approximate the fair price of European call option in subdiffusive B-S model.
\item[--] We have given condition under which the discrete scheme is stable and convergent.
\item[--] We have found the optimal choice of discretization parameter $\theta$ in dependence of subdiffusion parameter $\alpha$. Such numerical scheme is unconditionally stable, unconditionally convergent and has the lowest numerical error. 
\item[--] We have presented some numerical examples to illustrate introduced theory.
\end{itemize}
We believe that the numerical techniques presented in this paper can successfully be repeated for other fractional diffusion-type problems.
 \section*{Acknowledgments}
 The research of M.M. was partially supported by NCN-DFG Beethoven Grant No.2016/23/G/ST1/04083.



\bibliographystyle{spbasic}
\bibliography{Manuscript with changes}







\end{document}